*Commentary*

# From Materials Database to Materials Bank: Assetizing Data for AI-Driven Materials Innovation

Chenyao Ma [1], Di Zhang [2,3], Weibo Gong [4], Wei Du [1,4], Rui Su [1], Yuhang Chen [1], Kan Xu [1], Huan Gu [1], Limin Li [1,4,*], Piao Ma [1,4,*], Zhenghao Li [5,*] and Hao Li [2,*]

1 Suzhou MatSource Technology Co., Ltd., Suzhou 215000, China
2 Advanced Institute for Materials Research (WPI-AIMR), Tohoku University, Sendai 980-8577, Japan
3 Frontier Research Institute for Interdisciplinary Sciences (FRIS), Tohoku University, Sendai 980-8577, Japan
4 Gusu Laboratory of Materials, Suzhou 215000, China
5 State Key Laboratory of Advanced Environmental Technology, Department of Environmental Science and Engineering, University of Science and Technology of China, 230026, China
* Correspondence: lilimin@matsourceai.com (L.L.); mapiao@matsourceai.com (P.M.); lizhhao@ustc.edu.cn (Z.L.); li.hao.b8@tohoku.ac.jp (H.L.)

**Abstract:** Driven by high-throughput experimentation, computational modeling, and artificial intelligence (AI), materials data has expanded at an unprecedented rate. Conventional materials databases function only as passive repositories, archiving raw experimental records indiscriminately including both successful and failed data, without systematic value filtering or asset management. This creates a critical gap between massive data accumulation and actionable innovation, hindering the identification of high-potential materials and industrial translation. To address this bottleneck, we propose an industrialization-oriented Materials Bank, a dedicated value-filtering and assetization layer that operates beyond traditional databases. It does not merely curate high-quality data but systematically elevates qualified candidates into standardized, upgradable materials assets via a multi-dimensional BankCard framework covering scientific validity, synthesis feasibility, application readiness, and industrial value. By unifying databases, AI models, automated experimentation, and multi-criteria assessment into a cohesive closed-loop ecosystem, the Materials Bank establishes a clear trajectory from data to knowledge, candidate, asset, and product. It serves not as an enhanced database or screening tool, but as a decision infrastructure bridging academic discovery and industrial demand, offering a scalable paradigm to accelerate AI-driven materials innovation and deliver tangible real-world impact.

## 1. Introduction

Rapid advances in high-throughput experimentation, computation, and artificial intelligence (AI) have driven explosive growth in materials data. Traditional databases serve as essential repositories for raw data, including both successful and failed experimental records [1,2]. However, they remain confined to passive storage, lacking systematic evaluation and value-screening mechanisms. Consequently, databases cannot identify high-potential candidates or guide industrial translation, creating a persistent gap between data abundance and actionable materials innovation [3,4]. In response to this mismatch between massive raw data and industrial demand, a range of industrial material data platforms have been developed alongside the emerging concept of material assetization. Representative computational repositories such as Materials Project [5] and AFLOW [6] establish unified property screening criteria for computational bulk material candidates, while industrial informatics platforms including Granta EduPack introduce fixed multi-indicator scoring to assess commercial material performance. Nevertheless, most existing frameworks only adopt static evaluation metrics locked to preset weights and thresholds [7]. They lack full-lifecycle asset management coupled with automated experimental closed loops: for example, their scoring rules cannot shift priorities between fundamental computational screening of solid electrolytes and pilot-scale performance verification of electrocatalysts, failing to realize flexible multi-dimensional screening for diverse research scenarios.

To tackle this issue, we propose an industrialized *Materials Bank* tailored for AI-driven materials innovation. Built upon conventional databases, this AI-powered professional management framework centers on material assetization and industrial translation. Far from being a mere combination of curated databases, AI screening tools and TRL assessments, the Materials Bank relies on three core pillars: validated datasets, industrial maturity evaluation and closed-loop feedback. It integrates databases, AI predictive models, autonomous experimentation and multi-criteria evaluation into a unified innovation ecosystem focused on full-lifecycle asset governance. The system operates as a complete closed-loop workflow: raw database data feeds initial AI screening, prospective candidates are validated through automated experiments, and evaluation results are fed back to update both databases and the candidate inventory of the Materials Bank (Figure 1). High-value material assets are further advanced for industrial scaling. This workflow bridges fundamental research and industrial demands, forming a complete value progression chain: data → knowledge → material assets → end products.

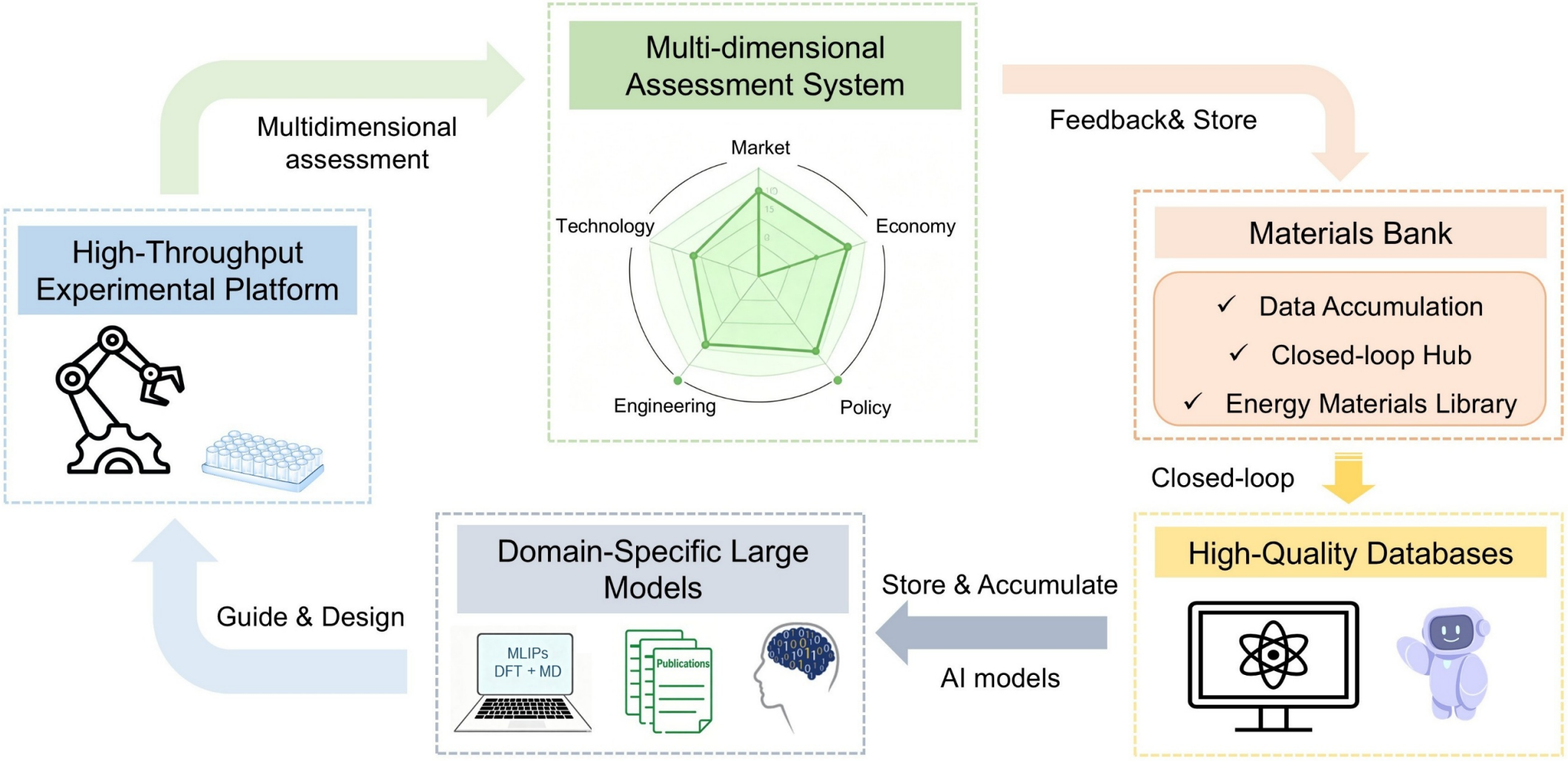

**Figure 1.** The closed-loop innovation ecosystem of the Materials Bank for AI-driven materials discovery. The Materials Bank acts as the core industrialization-oriented knowledge hub, underpinned by high-quality databases, domain-specific large models, modular high-throughput experimental platforms, and full-chain feedback mechanisms. This integrated ecosystem enables accelerated discovery, multi-dimensional evaluation, and iterative optimization of advanced energy materials.

(i) Data: The raw bottom-layer information collected by high-throughput experiments and computational simulations, including unfiltered successful and failed experimental records, crystal structural parameters, and primitive electrochemical signals. Stored passively in conventional materials databases without value screening or logical refinement, data serves as the fundamental raw input for the entire innovation pipeline.

(ii) Knowledge: Scientific laws, structure-property relationships and screening guidelines generalized from massive raw data via domain-specific large models. Knowledge extracts interpretable correlations out of fragmented datasets, completing the first value upgrade from scattered records to reusable scientific insights, which support rapid preliminary screening of promising material candidates.

(iii) Materials Assets: Standardized, iterable and upgradable material candidates that pass four-dimensional assessments (scientific validity, synthesis feasibility, application readiness, and industrial potential) and are formally managed via the Materials Bank's BankCard framework. Unlike theoretical knowledge alone, materials assets are experimentally validated, fully indexed, and treated as manageable long-term R&D resources with clear commercialization prospects, representing the transformation from theoretical insights to operable industrial R&D assets.

(iv) Products: Commercially viable materials or devices derived from high-grade materials assets after pilot-scale verification, mass-production optimization and market adaptation. While assets only embody industrialization potential at the laboratory stage, finished products realize tangible commercial value through full engineering scaling, marking the highest-value output of the whole closed-loop system.

The four tiers feature an inherent progressive dependency: data yields knowledge, which screens candidates standardized into formal material assets via multi-dimensional assessment; mature assets then undergo engineering iteration to create marketable products. Every upper tier refines and elevates the value of lower layers, outlining the full value chain spanning basic academic research to industrial implementation, and this tiered framework equips the Materials Bank with a scalable paradigm to expedite AI-led materials innovation and deliver tangible practical value.

## 2. Database

The rapid development of high-throughput experiments, computational simulations and AI has triggered explosive growth in materials data. As core data carriers, materials databases comprehensively collect various raw data covering both successful and failed experiments. They lay a solid foundation for intelligent research and experimental verification, provide sufficient data support for energy materials research, and consolidate the underlying basis of the entire materials innovation system [8,9]. Crucially, databases are repositories where only data accumulation and archiving are performed, while no value filtering, asset definition, or transformation management is conducted. At present, such data resources have been widely applied in the fields of solid electrolytes and electrocatalysis. Specifically, they facilitate the development of solid electrolytes with high ionic conductivity and the screening of high-performance catalysts, and effectively support mechanism analysis, performance optimization and rapid screening of candidate materials in relevant research directions.

## 3. Large Models

Supported by massive raw data stored in materials databases, domain-specific large models enable preliminary intelligent screening and prediction of materials. These models efficiently mine potential correlations within datasets and screen out promising candidate materials, greatly boosting screening efficiency. They build an effective link between basic data and candidate materials, delivering accurate guidance for subsequent experimental verification. Researchers established data-driven intelligent research frameworks based on large language models (LLMs) tailored for solid electrolytes, accelerating the development of innovative hydride-based solid electrolytes [10–12]. Advanced LLMs also serve as powerful tools for electrocatalytic materials research, clarifying inherent relationships among catalytic activity, structural stability, electronic configuration, defect morphology and reaction intermediates from abundant experimental data. Furthermore, multimodal and vision language models are capable of processing multi-source information including texts, graphs and crystal structures, realizing comprehensive extraction and integrated analysis of research data in materials science.

## 4. Modular Labs

Modular high-throughput experimental platforms conduct standardized verification on candidate materials screened by large models. They efficiently generate experimental data to confirm the scientific rationality and synthesis feasibility of target materials, laying a solid experimental foundation for further screening [13].

One typical example is one of the self-driving labs developed by Suzhou MatSource Technology. Equipped with high-energy ball milling, high-temperature sintering and solvothermal/hydrothermal reaction units, the self-developed platform enables fully automatic material preparation. After standardized post-treatment, samples are characterized in parallel via high-throughput XRD for structural analysis and multi-channel electrochemical tests for performance evaluation. All experimental data are stored in the Materials Bank as evidence for asset grading and status tracking, enabling traceable management and in-depth analysis (Figure 2).

Featuring flexible modular configuration, the platform adapts well to cathodes, anodes, solid electrolytes, and other energy material systems without adjusting the overall structure. Centralized intelligent control reduces manual work, shortens research cycles and minimizes human errors, enabling high-throughput parallel screening of diverse material formulas. Integrating automatic synthesis, efficient characterization and data management based on the Materials Bank, it builds a complete data-driven research loop and greatly improves research efficiency and experimental reliability.

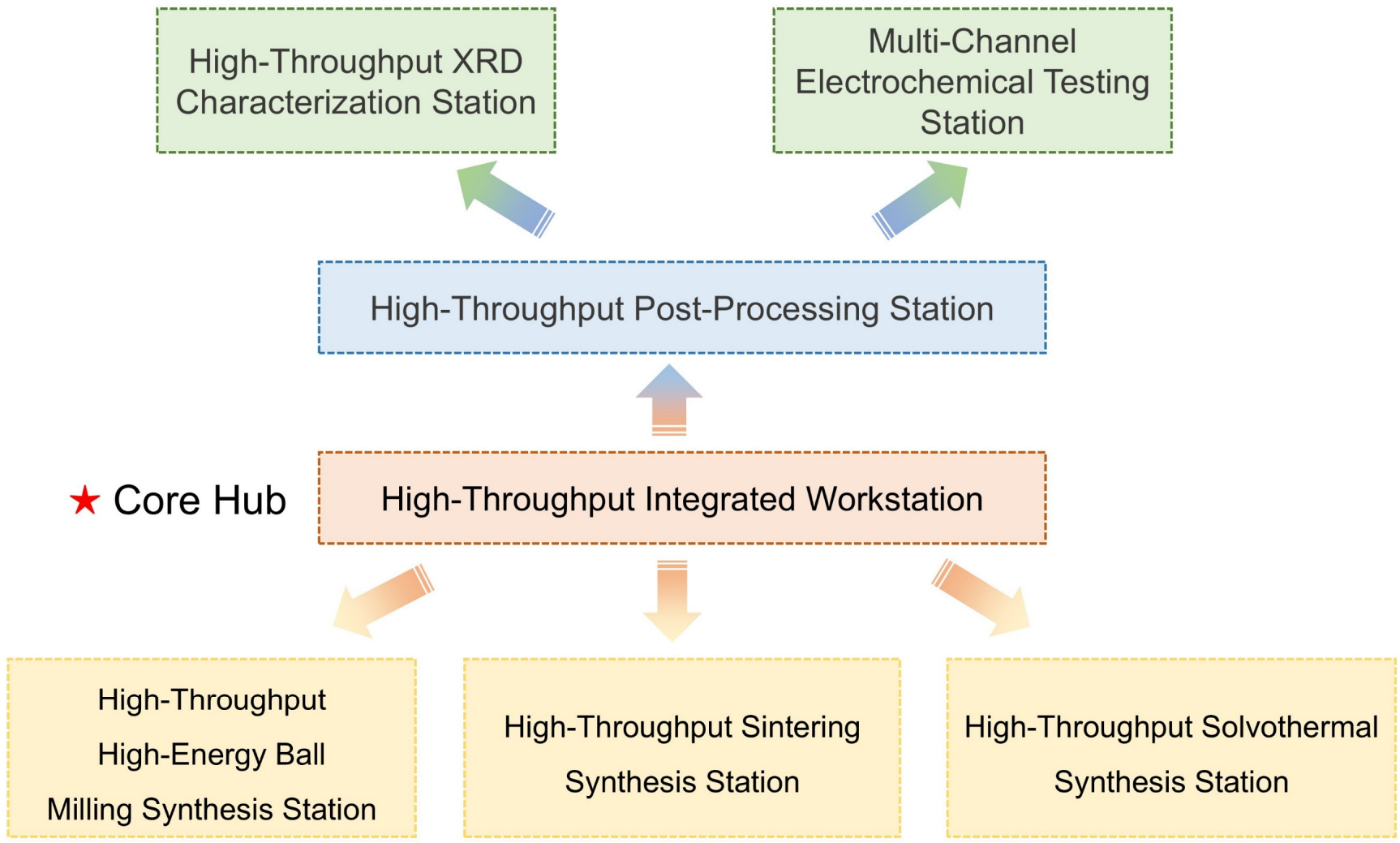

**Figure 2.** The red asterisk denotes the core hub, namely the High-Throughput Integrated Workstation that manages sample transfer and task scheduling across all synthesis, post-processing and characterization modules. It orchestrates three parallel synthesis stations (ball milling, sintering, solvothermal synthesis), delivers materials to the unified post-processing station, and implements automated characterization via high-throughput XRD and multi-channel electrochemical testing, establishing a closed-loop workflow for accelerated materials discovery and optimization.

## 5. Assessment

Modular automated high-throughput platforms enable parallel synthesis and preliminary characterization of candidate materials. Standardized data generated must pass a four-tier asset qualification system spanning from scientific discovery to industrial translation before being incorporated into the Materials Bank as formal assets. This system adopts progressive criteria: scientific validity, materials feasibility, application readiness, and industrial translation potential, with dynamically adjustable thresholds and weights to screen materials at each stage and avoid invalid data accumulation.

As a concrete illustration, two representative energy material research scenarios demonstrate the dynamic tuning logic of thresholds and weights. For fundamental screening of novel solid electrolyte candidates at the early laboratory stage, the framework assigns a high weight to scientific validity and raises the threshold for structural stability, while reducing the proportion of industrial translation potential; the primary goal is to eliminate materials with unreasonable crystal structures and poor theoretical ionic conductivity. In contrast, for electrocatalytic catalysts undergoing pilot trial development close to industrialization, the weight of synthesis scalability, element abundance and production cost is greatly elevated, together with tighter thresholds for cycling durability and batch reproducibility, so as to sift out candidates compatible with mass manufacturing. Such differentiated parameter settings explicitly reflect the flexible adjustment strategy of the four-tier assessment system across distinct research contexts.

Tiered data flows back dynamically: early-stage data optimizes large model prediction accuracy, while mature data iteratively refines evaluation criteria, forming a self-reinforcing loop. Notably, this multi-dimensional assessment is not merely an extended TRL evaluation; it serves as the core grading mechanism for material assets [14]. The Materials Bank thus functions as a tiered asset library covering the full R&D pipeline rather than a mere industrial database or assessment tool, effectively bridging lab research and translation, and improving the reliability, efficiency and translation potential of energy materials R&D. This four-dimensional scoring standard forms the essential dividing line between curated raw datasets and formal material assets archived in the Materials Bank. Curated data only undergoes basic cleaning and formatting without multi-dimensional industrial evaluation, while material assets are curated data that pass all four evaluation metrics and are managed as standardized, lifecycle-tracked R&D resources.

## 6. Materials Bank

Table 1 lists seven core functional modules of the BankCard framework, which jointly enable full-lifecycle evaluation of materials from basic research to industrialization. The Identity module defines unique material fundamental information; the Evidence module records data sources and their reliability uncertainty; the Performance module collects key functional indicators of energy materials; the Feasibility module assesses synthesis and scale-up constraints; the Readiness module quantifies industrial maturity based on TRL, compatibility and reproducibility; the Value module measures economic, patent and market potential; and the Status module marks the current R&D progress of each material asset. The multi-angle indicators covered by all seven modules complement each other without omission, completely matching the multi-stage screening demands of solid electrolytes, electrocatalysts and other energy materials. This modular design forms the core asset classification logic of BankCard, supporting flexible dynamic evaluation for diverse research scenarios.

As the core assetization and transformation hub of AI-driven material innovation systems, the Materials Bank does not merely collect data but governs materials as upgradable assets. It sorts out scattered original data, candidate materials and experimental achievements generated in previous research stages, and converts them into standardized, reusable and iterable high-quality material assets. It formulates unified BankCard asset admission rules to standardize the warehousing process and set practical operational specifications. Only materials that pass multi-dimensional evaluation and meet all specified access indicators can be officially included to realize standardized management of material assets [15]. In addition, to safeguard asset integrity and long-term data reliability, the Materials Bank implements full traceability and multi-version archiving for all stored assets. All raw datasets, initial four-dimensional assessment records and subsequent experimental/computational logs are permanently retained without overwriting or removal. If new measurements or simulations contradict an asset's documented properties, the platform creates a labeled new asset version and launches a secondary re-evaluation via the four-dimensional scoring framework. Conflicting outcomes are retained to retrain domain-specific large models instead of being discarded as defective data; these discrepancies act as informative training samples to improve screening precision, aligning with the insight that experimental inconsistencies facilitate iterative progress for both researchers and AI models.

**Table 1.** The "BankCard" framework establishes standardized, multi-dimensional admission and classification criteria for materials entering the Materials Bank. It includes modules for material identity, evidence reliability, performance metrics, synthesis feasibility, technological readiness, application value, and development status, ensuring only verified, high-potential materials are archived and tracked throughout the innovation pipeline.

| Module | Content |
| --- | --- |
| Identity | composition, structure, phase, synthesis route |
| Evidence | literature source, experiment, computation, uncertainty |
| Performance | activity, conductivity, stability, selectivity, etc. |
| Feasibility | synthesis difficulty, scalability, element abundance |
| Readiness | TRL, device compatibility, reproducibility |
| Value | cost, IP space, sustainability, market/application fit |
| Status | archived, candidate, validated, scalable, deployable |

## 7. Discussion and Summary

The proposed Materials Bank is not merely a curated database combined with AI screening and TRL assessment. Unlike conventional databases that passively store raw data, it functions as a dedicated assetization layer built on value filtering and systematic evaluation. Its core innovation lies in the BankCard framework, which standardizes multi-dimensional asset grading and lifecycle management, transforming fragmented data into reusable materials assets. By integrating data curation, AI prediction, automated experimentation, and closed-loop feedback, it bridges academic discovery and industrial demand. In the future, when industry requires materials with specific new properties, it can turn to the Materials Bank just as people seek funds from banks, enabling fast and precise matching of qualified candidates. Further optimization of evaluation criteria and AI-experimental integration will accelerate advanced energy materials development and industrialization, driving next-generation energy technologies.

## Author Contributions

C.M.: conceptualization, methodology, writing—original draft preparation, literature curation, formal analysis; D.Z.: literature curation, writing—reviewing and editing, validation; W.G.: formal analysis, literature curation; W.D.: resources, writing—reviewing and editing; R.S.: visualization, formal analysis; Y.C.: literature curation, visualization; K.X.: formal analysis, literature curation; H.G.: resources, literature curation; L.L.: supervision, writing—reviewing and editing, funding acquisition; P.M.: supervision, conceptualization, writing—reviewing and editing; Z.L.: supervision, conceptualization, validation; H.L.: supervision, conceptualization, project administration, writing—reviewing and editing. All authors have read and agreed to the published version of the manuscript.

## Funding

This research was financially supported by Suzhou MatSource Technology Co., Ltd. and Gusu Laboratory of Materials under Grant No. Y2501.

## Conflicts of Interest

Authors Chenyao Ma, Wei Du, Rui Su, Yuhang Chen, Kan Xu, Huan Gu, Limin Li and Piao Ma are employed by the company, Suzhou MatSource Technology Co., Ltd. The company did not exercise control over the study design, data collection, data analysis, interpretation of results, manuscript preparation, or the decision to submit the article for publication. Given the role as co Editor-in-Chief of *AI for Materials*, Hao Li had no involvement in the peer review of this paper and had no access to information regarding its peer-review process. Full responsibility for the editorial process of this paper was delegated to another editor of the journal. The remaining authors declare no conflicts of interests.

## Use of AI and AI-Assisted Technologies

No AI tools were utilized for this paper.